\documentstyle[psfig,epsf,prd,aps,floats,12pt]{revtex}  
\def\be{\begin{equation}} 
\def\ee{\end{equation}} 
\def\ba{\begin{eqnarray}} 
\def\ea{\end{eqnarray}}

\def\al{\alpha}

\def\be{\begin{equation}}
\def\ee{\end{equation}}
\def\ba{\begin{eqnarray}}
\def\ea{\end{eqnarray}}

\begin{document} 

\title{Generating non-Gaussian maps with a given power spectrum
and bispectrum}
\author{Carlo R. Contaldi$^1$ and Jo\~ao Magueijo$^3$\\
$^1$ Canadian Institute for Theoretical Astrophysics, 60 St. George
Street, Toronto, ON M5S
3H8, CANADA\\
$^2$ Theoretical Physics, The Blackett Laboratory, 
Imperial College, Prince Consort Rd., London, SW7 2BZ, U.K.} 
\date{\today}

\maketitle 
 
\begin{abstract} 
We propose two methods for generating non-Gaussian maps with fixed
power spectrum and bispectrum. The first makes use of a recently proposed 
rigorous, non-perturbative, Bayesian  framework for generating non-Gaussian
distributions. The second uses a simple superposition of Gaussian
distributions. The former is best suited for generating mildly 
non-Gaussian maps, and we discuss in detail the limitations of 
this method. The latter is better suited for the opposite situation, 
i.e. generating strongly non-Gaussian maps. The ensembles produced 
are isotropic and the power
spectrum can be jointly fixed; however we cannot set to
zero all other higher order cumulants (an unavoidable mathematical
obstruction). We briefly quantify the 
leakage into higher order moments present in our method. We finally present 
an implementation of our code within the HEALPIX package \cite{healpix}. 
\end{abstract}

\section{Introduction}
In most inflationary scenarios the fluctuations in the matter fields
generated by the oscillating inflaton display Gaussian statistics. This
requires the temperature fluctuations in the Cosmic Microwave 
Background (CMB) to be Gaussian distributed to a very high degree of
accuracy at sufficiently large angular scales. On smaller scales the
effect of late time non-linear evolution will
introduce a certain amount of non-Gaussianity in the CMB. 
The study of the non-Gaussianity of CMB fluctuations is therefore
crucial to both the understanding of the fundamental processes
generating the fluctuations and to the understanding of the
various foreground and astrophysical contributions. 

In addressing an issue of this nature with reference to observational
strategies it is important to be able to simulate CMB maps 
with non-Gaussian signatures. These can then be used in the
refinement of estimation techniques and the design of evermore
accurate satellite, balloon-borne, and ground-based experiments.
It seems therefore desirable to develop fast algorithms for
simulating not only Gaussian signals (as extensively done in the 
past, e.g.\cite{healpix}), 
but also maps allowing for non-Gaussianity. In the past the 
bispectrum has proved an invaluable tool in studying CMB non-Gaussianity 
(see for instance \cite{fmg,mag1,haav,sper,kog,Rocha}). Also 
algorithms generating Gaussian maps usually use the power spectrum
as a controlling parameter. We therefore seek to complement 
earlier work by producing 
an algorithm for generating non-Gaussian maps with fixed power spectrum 
and bispectrum. As a matter of fact the algorithm we shall propose
can be extended to fix simultaneously  any higher order moment,
naturally at a computational cost.

The difficulty in generating non-Gaussian maps is in part due to the
lack of suitable Probability Distribution Functions (PDFs) with the
required parametrisation, and in part due to the requirement that
non-zero higher moments require the multivariate generation of
correlated sets of modes. The latter is imposed by statistical isotropy,
in all cases except in Gaussian maps. We shall address
the first of these problems with reference to the recent work in \cite{Rocha},
in which a rigorous, non-perturbative, Bayesian  framework for generating 
non-Gaussian distributions was proposed. We also propose an alternative
PDF, a simple superposition of Gaussian
distributions. We then address the second problem with a rather
simple trick for creating the required mode correlations enforcing
isotropy. 

This paper is organised as follows. In section~\ref{meth} we introduce
two exact, non-perturbative univariate distributions which produce
non-Gaussian ensembles with fixed variance and skewness. In
section~\ref{higher} we derive the higher moments for both
distributions. Section~\ref{making} shows how we can employ the
non-Gaussian 1-D distributions to generate isotropic ensembles of
non-Gaussian maps with given angular power spectra and
bispectra. We show examples of such distributions in
section~\ref{results} and discuss the results and applications of this
work section~\ref{disc}.

\section{Exact non-Gaussian 1-D distributions}\label{meth}

In this section we introduce two classes of distributions which can 
be employed to generate random numbers with exactly specified second and third
moments. The first was introduced as a non-perturbative,
non-Gaussian
likelihood in the Bayesian analysis of the CAT/VSA data
\cite{Rocha}.  It was originally introduced, in a theoretical
context, in the study of off-the-ground state perturbations
in the inflaton field\cite{Contaldi}. 
The second class of distributions is an {\it ad hoc} solution obtained by
requiring the simplest superposition of Gaussian PDFs that results in a
distribution with a given second and third moments.

Both PDF functions can be used to generate distributions with fixed power
spectra and bispectra (and optionally higher order spectra). 
The first is best suited for mild non-Gaussianity; the latter
for strong departures from Gaussianity. 

\subsection{The non-perturbative harmonic likelihood}

We now summarise the method of Rocha et al. \cite{Rocha,Wave}.
Let $x$ represent a general random variable.  We build its
distribution from the space of wave-functions which are 
energy eigenmodes of a linear harmonic oscillator (see e.g
\cite{qm}). We have that
\be\label{psiexp} 
\psi(x)=\sum \alpha_n\psi_n (x),
\ee 
where $n$ labels the energy spectrum $E_n=\hbar \omega (n+1/2)$,
and 
\be 
\psi_n(x)=C_nH_n{\left(x\over 
{\sqrt 2} \sigma_0\right)}e^{-{x^2\over 4 \sigma_0^2}}, 
\ee 
with normalisation fixing $C_n$ as 
\be 
C_n={1\over (2^n n!{\sqrt{2\pi}}\sigma_0)^{1/2} }\, . 
\ee 
The only constraint upon the amplitudes $\alpha_n$ is
\be\label{const}     
\sum |\alpha_n|^2 =1,
\ee
which can be eliminated explicitly by imposing the condition
\be\label{elim}
\alpha_0= \sqrt{1- \sum_1^\infty |\alpha_n|^2 }.
\ee
$\sigma_0^2$ is the variance  associated with the (Gaussian) probability 
distribution for the ground state $|\psi_0|^2$.  We define Hermite 
polynomials $H_n(x)$ as  
\be\label{herm} 
	H_n(x) = (-1)^n e^{x^2} \frac{d^x}{d x^n}e^{- x^2}, 
\ee 
with normalisation  
\be\label{ortho} 
\int^\infty_{-\infty} e^{-x^2}H_n(x)H_m(x)dx=2^n \pi^{1/2} n!\delta_{nm}. 
\ee 
The most general probability density is thus:
\be\label{density}
P=|\psi|^2={e^{-{x^2\over 2 \sigma_0^2}}\over {\sqrt{2\pi}}}
|\alpha_n C_n H_n{\left(x\over {\sqrt 2}\sigma_0\right)}|^2
\ee
The ground state (or ``zero-point'') fluctuations are Gaussian, 
but any admixture with other states will be reflected 
in a non-Gaussian distribution function. 

It can be shown \cite{Wave,Rocha} that the $\alpha_n$ reduce 
to the cumulants $\kappa_n$ (up to a multiplicative constant) 
for mild, ``perturbative'', non-Gaussianity. This is achieved by reducing
the probability density (\ref{density}) to an asymptotic expansion
around the Gaussian distribution function (the so-called 
Edgeworth expansion). Such asymptotic expansions
on the space of orthonormal Hermite polynomials suffer from the fact
that truncations at a finite order of cumulants lead to
pseudo-distributions which are not positive definite. This is not the case
here and the advantage of using the $\alpha_n$ over cumulants
is that setting all but a finite number of them
to zero still leads to proper distributions. In some sense the $\alpha_n$
are non-perturbative generalizations of cumulants. 

The above probability density may be easily applied to
generate a centered distribution with a fixed variance and
skewness. Let us start with the PDF
\be
	P(x) = \frac{\exp(-x^2/2\sigma^2)}{\sqrt{2\pi}\sigma}\left[
\al_0+\frac{\al_1}{\sqrt{2}}H_1\left(\frac{x}{\sqrt{2}\sigma}\right)+\frac{\al_2}{\sqrt{8}}H_2\left(\frac{x}{\sqrt{2}\sigma}\right)
+\frac{\al_3}{\sqrt{48}}H_3\left(\frac{x}{\sqrt{2}\sigma}\right)\right]^2,
\ee
in which all the $\alpha_n$ are real and we have set all $\alpha_n =
0$ for $n>3$. We then calculate moments around
the origin defined as 
\be
	\mu_n = \langle x^n \rangle = \int^{\infty}_{-\infty} x^nP(x)dx.
\ee 
For a normalised density we have $\mu_0 =
\al_0^2+\al_1^2+\al_2^2+\al_3^2=1$  and the first three moments given by
\ba
	\mu_1 &=&
(2\sigma^2)^{1/2}\left(2\al_1\al_2+\sqrt{2}\al_0\al_1+\sqrt{6}\al_2\al_3
\right),\nonumber\\
	\mu_2 &=&
\sigma^2\left(\al_0^2+3\al_1^2 +
5\al_2^2+7\al_3^2+2\sqrt{2}\al_0\al_2+2\sqrt{6}\al_1\al_3\right),\nonumber\\
	\mu_3
&=&(2\sigma^2)^{3/2}\left(\frac{3}{\sqrt{2}}\al_0\al_1+6\al_1\al_2+\frac{9\sqrt{3}}{\sqrt{2}}\al_2\al_3+\sqrt{3}\al_0\al_3
\right) .
\ea
The aim is to have a centered distribution with $\mu_1=0$. This can be
achieved by setting  $\al_1=0$ and $\al_2=0$ in which case we are left
with the system
\ba\label{sys}
1 &=&\al_0^2+\al_3^2,\nonumber\\
\mu_2 &=&
{\sigma^2}\left(\al_0^2+7\al_3^2\right),\nonumber\\
\mu_3 &=&(2\sigma^2)^{3/2}\sqrt{3}\al_0\al_3.
\ea
Setting $\alpha_0=\sqrt{ 1-\alpha_3^2}$, we then solve for  $\sigma$
and $\al_3$ for particular values of the required second and third
moments $\mu_2$ and $\mu_3$
\ba
	\sigma^3 &=&
\frac{\mu_3}{2\sqrt{6}(1-\al_3^2)^{1/2}\al_3},\nonumber\\
	\al_3^2 &=& \frac{1}{6}\left(\frac{\mu_2}{\sigma^2}-1\right). 
\ea 
a problem we deal with numerically. Hence we arrive at a 1-D
centered distribution with fixed variance and skewness.

By restricting ourselves to only two parameters we have in fact
constrained the space of possible distribution functions. This is
reflected in the fact that we cannot generate distributions with any
given variance and skewness. By studying the function for the relative
skewness $s$
\be
s \equiv {\mu_3\over \mu_2^{3/2}}={2\sqrt{6}(1-\alpha_3^2)\alpha_3
\over (1+6\alpha_3^2)^{3/2}}
\ee
it is easy to see that the maximal relative skewness is 
\be
s = \pm 0.739 \ \ \ \ \mbox{at} \ \ \ \  \alpha_3 = \pm 0.278.
\ee
In general our method can generate higher values of $s$ (since it
can generate any distribution), but for that purpose one needs
more parameters $\alpha_n$. The generalization of the above construction
for more parameters is trivial if somewhat tedious.

\begin{figure}
\centerline{\hbox{\psfig{figure=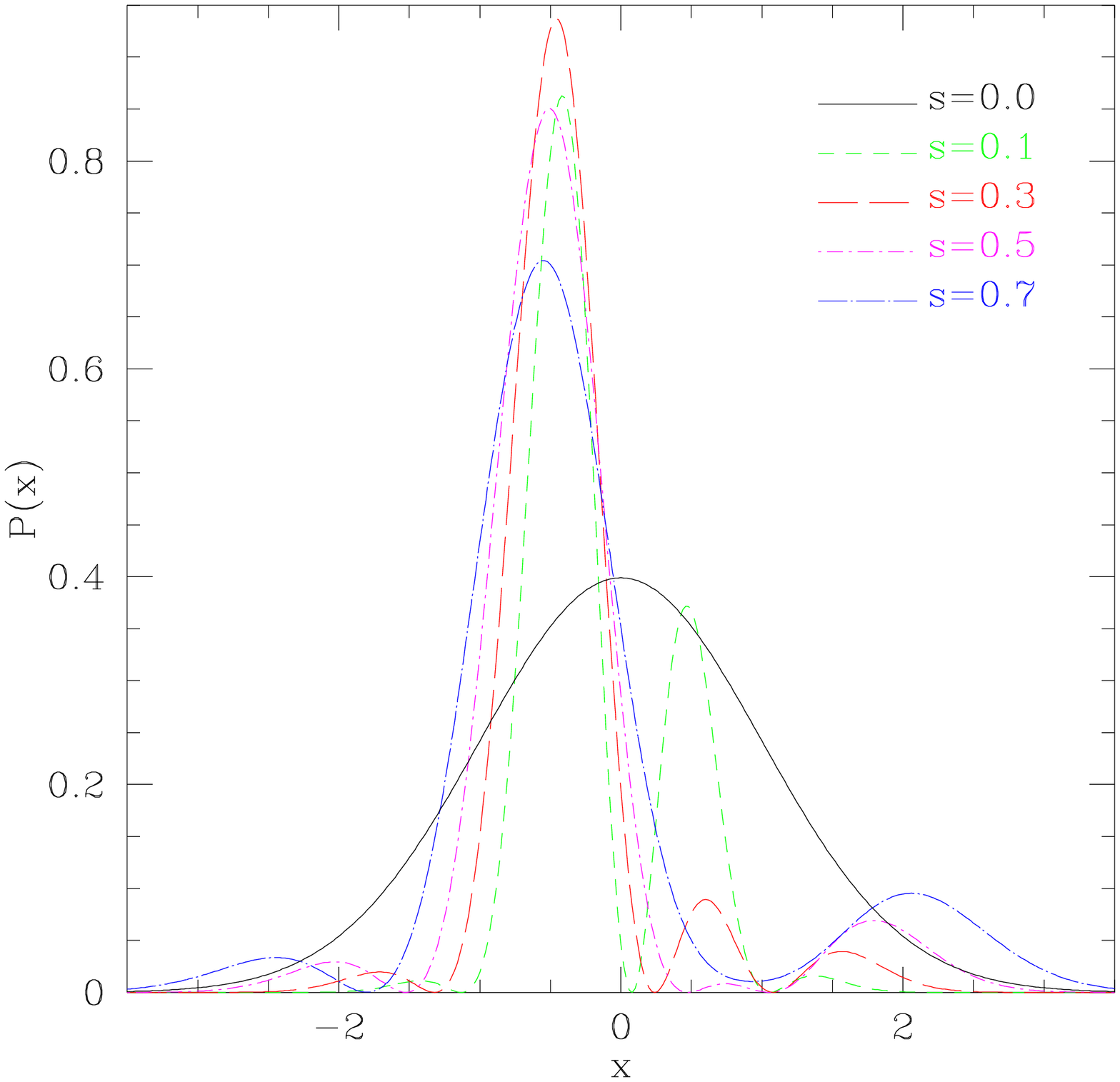,width=9cm}
\psfig{figure=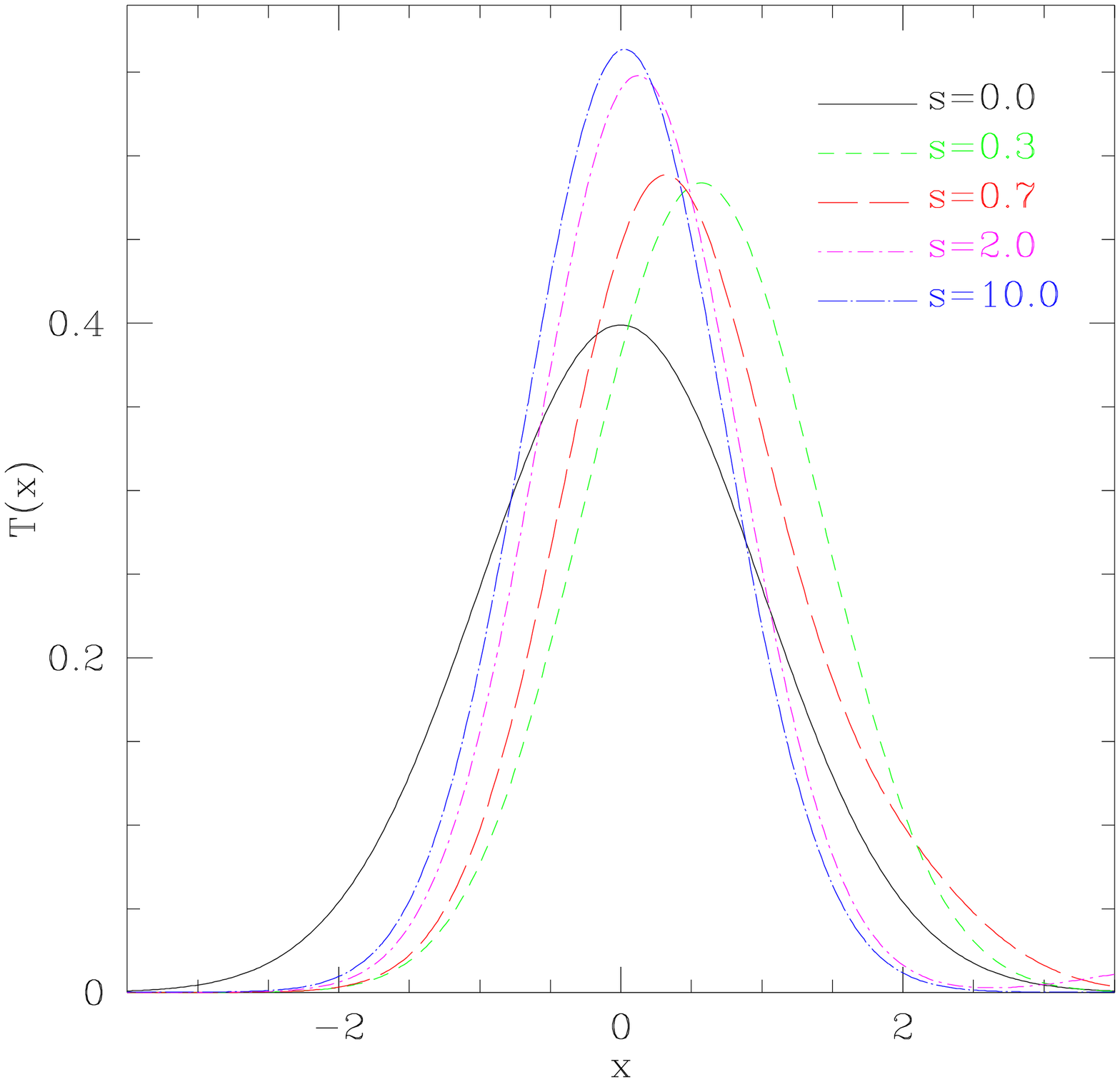,width=9cm} }}
\caption{The left panel shows the distribution $P(x)$ for various
values for
the relative skewness parameter $s=\mu_3/\mu_2^{3/2}$ ($0.0\le s \le
0.7$) . The 
right panel shows 
the distribution $T(x)$ for $0.0\le s\le10.0$.}
\label{px}
\end{figure}

\subsection{A tailor made distribution function}

One way to overcome the limitations of the distribution described
above without increasing the number of parameters 
is to build a distribution by `summing' Gaussians. In fact it is
obvious that the
distribution $P(x)$ can be well approximated by a series of uncentered
Gaussian functions of different heights and widths. Consider, for example, the
distribution obtained by the superposition of three
Gaussian distributions 
\be
	T(x) =  \frac{1}{(2+\beta_0)\sqrt{\pi}}\left( e^{-x^2}+e^{-(x-\beta_0\beta_1)^2}+\beta_0e^{-(x+\beta_1)^2}\right).
\ee
In this case the first three moments of the distribution are
given by
\ba\label{sys1}
	\mu_1 &=& 0,\nonumber\\
	\mu_2 &=& \frac{1}{2(2+\beta_0)}\left( 2+\beta_0+2\beta_0\beta_1^2(1+\beta_0)\right),\nonumber\\
	\mu_3 &=& \frac{\beta_0\beta_1^3(\beta_0^2-1)}{2+\beta_0}.
\ea
In contrast to the previous distribution the value of $s$ is now
effectively  unconstrained. In practice we find that in this case the
range of $s$ is limited only by numerical effects in the random number
generation process. 

Fig.~\ref{px} compares the distributions $P(x)$ and $T(x)$ for various
values of the relative skewness $s$. The
former may be considered somewhat `unphysical' since the minima in the
functions will tend to produce `holes' in the distribution of
variates for each particular set of moments (this is not surprising since the
PDF originates from the wavefunction of an oscillator). We should point out
though that the present exercise is aimed at producing maps for the
sole purpose of testing statistical tools and not to reproduce
theoretically motivated models of the cosmic microwave
backgrounds. The adaptation of the method to more suitably physical
non-Gaussian distribution functions such as the PDF $T(x)$ above is
trivial as we have shown. 

A word of caution is in order, concerning the numerical aspects
of generating univariate random numbers with a given non-Gaussian
PDF. The Jacobian required to generate the random variables using the
transformation method is not easily computable in the case of the
PDFs $P(x)$ and $T(x)$. We advocate, instead,  
the use of the well known rejection method
\cite{Press} to generate the required one dimensional random
variables. 


\section{Leakage into the higher moments}\label{higher}

Both distributions have general solutions for all higher order
moments $\mu_n$. The fourth and sixth order moments are of particular
interest since they generate the sample variance and hence the cosmic
variance in the observed power spectrum and bispectrum of the realisations. For
$n$ even we find the following 
relatively simple expressions for the moments 
\be\label{mom1}
	\mu_n=\frac{(2\sigma)^n}{\sqrt{\pi}}\left[\al_0^2\Gamma\left(\frac{1+n}{2}\right)+\al_3^2\left\{ 3\Gamma\left(\frac{3+n}{2}\right)-4\Gamma\left(\frac{5+n}{2}\right)+\frac{4}{3}\Gamma\left(\frac{7+n}{2}\right)\right\} \right],
\ee
for the distribution $P(x)$ and
\be\label{mom2}
	\mu_n =
\frac{\Gamma\left(\frac{1+n}{2}\right) }{(2+\beta_0)\sqrt{\pi}}\left[e^{-\beta_1^2}\beta_0
M\left(\frac{1+n}{2},\frac{1}{2},\beta_1^2\right)+e^{-\beta_0^2\beta_1^2} M\left(\frac{1+n}{2},\frac{1}{2},\beta_0^2\beta_1^2\right)+1\right],
\ee
for the distribution $T(x)$ where $M(a,b,c)$ are the confluent
hypergeometric functions \cite{Abramowitz}. Unfortunately the
non-linear dependence of the parameters $\sigma,\al_n$ and $\beta_n$
on the required ensemble moments complicates the analytical form for
the sample variances $\sigma^2(\mu_2)$ and $\sigma^2(\mu_3)$. In
practice therefore the sample variances would be calculated
numerically via the above expressions and making use of the solutions
to the systems (\ref{sys}) and (\ref{sys1}) or via Monte-Carlo simulations.

\section{Generating non-Gaussian maps with fixed angular power spectrum
and bispectrum}\label{making}
We now propose to apply this method to the generation of non-Gaussian 
maps with fixed angular power spectrum $C_\ell$ 
and bispectrum $B_\ell$. The broader context is
the generation of non-Gaussian signals to be be used in 
simulations of upcoming satellite experiments. 

The distributions of the previous section have the disadvantage that
they can only 
generate non-Gaussian 1-D distributions. As a result, when a
likelihood constructed using $P(x)$ is applied 
to CMB maps as in Rocha et al., it probes a combination of
non-Gaussianity {\it and} anisotropy. 
Consider, for instance, the generation of a full-sky map with
temperature fluctuations 
$\frac{\Delta T}{T}({\bf n})$, by means of its 
spherical harmonic components:
\begin{eqnarray} 
\frac{\Delta T}{T}({\bf n})=\sum_{\ell m}a_{\ell m}Y_{\ell m}({\bf n}). 
\label{almdef} 
\end{eqnarray} 
If we set $\alpha_3\neq 0$ in the above distribution and use it to
generate a set of $\Re (a_{\ell 0})$ we will have  
$\langle(\Re a_{\ell 0})^3\rangle\neq 0$, with $\langle (\Re a_{\ell m})^3\rangle= 0$ for $m\neq 0$.
Isotropy, on the other hand, imposes ``selection rules'' upon
correlators, in this case
\begin{eqnarray}
{\langle a_{\ell_1 m_1}a_{\ell_2 m_2} a_{\ell_3 m_3}\rangle}=
\left ( \begin{array}{ccc} \ell_1 & \ell_2 & \ell_3 \\ m_1 & m_2 & m_3
\end{array} \right ) B_{\ell_1\ell_2\ell_3},
\label{defnpoint}
\end{eqnarray}
where the quantity $(\ldots)$ is the Wigner $3J$ symbol, and 
the coefficients
$B_{\ell_1\ell_2\ell_3}$ are the bispectrum (abbreviated to $B_{\ell}$
for the case $\ell_1=\ell_2=\ell_3$). 
Hence the distribution we have just generated is not only non-Gaussian
but also automatically anisotropic since all third order correlators
except for $\langle a_{\ell 0}\rangle$ are zero.

If we are to impose isotropy, we must necessarily have correlated 
$a_{\ell m}$, a feature not allowed by the method of Rocha et al. In
the present context a correlated set of $a_{\ell m}$ is
equivalent to a series of coupled harmonic oscillators.
The obvious way to achieve the necessary correlations would be to use
the Hilbert space of coupled harmonic oscillators to set up 
the most general multi-variate distribution but this proves to be impractical.
Here we introduce a simple but crucial modification which restores
isotropy. 

We propose to set up an isotropic ensemble in two steps. First 
we generate an anisotropic ensemble by drawing all $a_{\ell m}$ 
from a Gaussian distribution with variance spectrum $C_\ell$, 
except for the $m=0$ mode. The latter is given one of the PDFs
described in the previous section, with  variance $C_\ell$
and skewness 
\be
s_\ell= \left 
( \begin{array}{ccc} \ell & \ell & \ell  \\ 0 & 0 & 0
\end{array} \right )^{-1} B_{\ell} 
\ee
The appropriate PDF is enforced using the rejection method,
as described above.  For each set of $C_{\ell}$ and $B_{\ell}$ we solve the
systems given by Eqs.~(\ref{sys}) or (\ref{sys1}) numerically to obtained
the required values for the parameters
$\sigma$, $\alpha_0$ and $\alpha_3$ or $\beta_0$ and $\beta_1$ . We
label the resulting ensemble 
$a$, for anisotropic. 

We then apply a random rotation to each realization in the ensemble $a$,
with a uniform distribution of Euler angles. In this way we 
arrive at an isotropic ensemble of temperature maps, labelled by $i$,
with spherical harmonic coefficients
\be
	b^{\ell}_m = \sum_{m'} {\cal D}^{\ell}_{mm'}(\Omega) a^{\ell}_{m'}
\ee
where ${\cal D}^{\ell}_{mm'}$ is the rotation matrix, and $\Omega$
denote the 3 Euler angles. The 2-point correlators are
\be
{\langle b_{\ell_1m_1} b^\star _{\ell_2 m_2}\rangle}_i
= \sum_{m_1' m_2'}
{\langle {\cal D}^{\ell_1}_{m_1m_1'}(\Omega)
{\cal D}^{\ell_2\star}_{m_2m_2'}(\Omega)\rangle}_\Omega
{\langle a_{\ell_1m_1'} a^\star _{\ell_2 m_2'}\rangle}_a
=\delta_{\ell_1\ell_2}\delta_{m_1m_2}C_{\ell_1},
\ee
hence we have $C_{\ell}^i \equiv C_{\ell}^a$.

The 3-point correlators are zero for any correlator involving
different $\ell$. However we now have
\ba
{\langle b_{\ell m_1} b_{\ell m_2} b_{\ell m_3}\rangle}_i
&=& \sum_{m_1' m_2' m_3'}
{\langle {\cal D}^{\ell}_{m_1m_1'}(\Omega)
{\cal D}^{\ell}_{m_2m_2'}(\Omega){\cal D}^{\ell}_{m_3m_3'}(\Omega)
\rangle}_\Omega
{\langle a_{\ell m_1'} a_{\ell m_2'} a_{\ell m_3'}\rangle}_a
\nonumber \\
&= & {\langle {\cal D}^{\ell}_{m_1 0}(\Omega)
{\cal D}^{\ell}_{m_2 0}(\Omega){\cal D}^{\ell}_{m_3 0}(\Omega)
\rangle}_\Omega \times s_\ell\nonumber \\
& =& {\langle Y^{\ell}_{m_1}(\Omega)
Y^{\ell}_{m_2}(\Omega) Y^{\ell}_{m_3}(\Omega)
\rangle}_\Omega \times s_\ell\nonumber \\
&=&
{\left ( \begin{array}{ccc} \ell & \ell & \ell  \\ 0 & 0 & 0
\end{array} \right )} {\left( \begin{array}{ccc} \ell & \ell & \ell  \\ 
m_1 & m_2 & m_3
\end{array} \right )} s_\ell\nonumber \\
&=& {\left( \begin{array}{ccc} \ell & \ell & \ell  \\ 
m_1 & m_2 & m_3
\end{array} \right )} B_\ell
\ea
(note that with one index $m$ set to zero the rotation matrices
reduce to spherical harmonics, one of the Euler angles becoming
irrelevant). Again we have recovered the original bispectrum with
$B_{\ell}^i \equiv B_{\ell}^a$.

Hence the procedure we have defined produces the desired isotropic
ensemble with a fixed angular power spectrum and bispectrum.
The random rotations produce the necessary correlations between
the $a_{\ell m}$ coefficients to ensure isotropy. However, by means
of this procedure, we are unable to generate maps with non vanishing
inter-$\ell$ bispectrum coefficients (as studied in 
\cite{mag1,haav,sper,kog}).

Note that we can also draw {\it all} $a_{\ell m}$ from the 1-D PDFs
introduced above, and then subject them to a random rotation. The
argument presented here still goes through, and an isotropic ensemble
is still obtained. However, as we shall see later, the cosmic variance
in the estimators for the $C_{\ell}$ and $B_{\ell}$ will be larger in this
case.

\section{Simulated non-Gaussian maps}\label{results}

We now present CMB temperature maps using the prescriptions detailed
in the previous sections.  We generate full sky maps
pixelised using
the HEALPIX \cite{healpix} package at pixelisation levels $128$ and
$512$, equivalent to $196,608$ and $3,145,728$ pixels respectively. We
generate maps which include a cosmological signal in the form of a
standard $\Lambda$CDM power spectrum computed using CMBFAST
\cite{cmbfast} and finite beam sizes. Instrument specific noise and
physically motivated profiles for the bispectrum will be treated in
further work \cite{prep}.

For simplicity we shall assume a bispectrum which is fixed
relative to the power spectrum (i.e. $B_{\ell} = sC_{\ell}^{3/2}$)
with a ratio which is the same for all scales (i.e. with $s_{\ell}\equiv s$
constant). This is a rather artificial asumption since in practice the
host of non-linear, secondary and foreground contributions to the CMB
\cite{bisp} and possible primordial non-Gaussianity will arise at different
angular scales. Theoretically motivated bispectra and the
addition of instrument specific noise will be discussed 
in further work \cite{prep}, and are straightforward to obtain with 
our method at no extra computational cost.
We use both the PDFs discussed above to generate the 1-D
distributions required for the production of the initial anisotropic
ensembles.

\begin{figure}
\centerline{\psfig{file=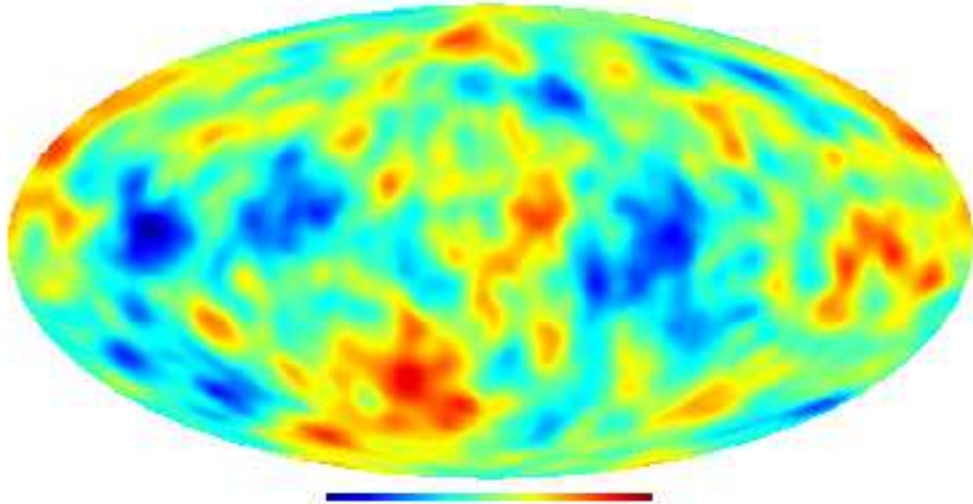,width=14 cm,angle=90}}
\caption{HEALPIX map of a noiseless, Gaussian CMB realisation at COBE
beam resolution (normalised temperature units). The map 
was generated using a standard $\Lambda$CDM power spectrum ($\Omega_{\Lambda} = 0.7$, $\Omega_{CDM} = 0.25$,
$\Omega_{b} = 0.05$, $h=0.7$ and $n_s = 1.0$).}
\label{COBE}
\end{figure}

\begin{figure}
\centerline{\psfig{file=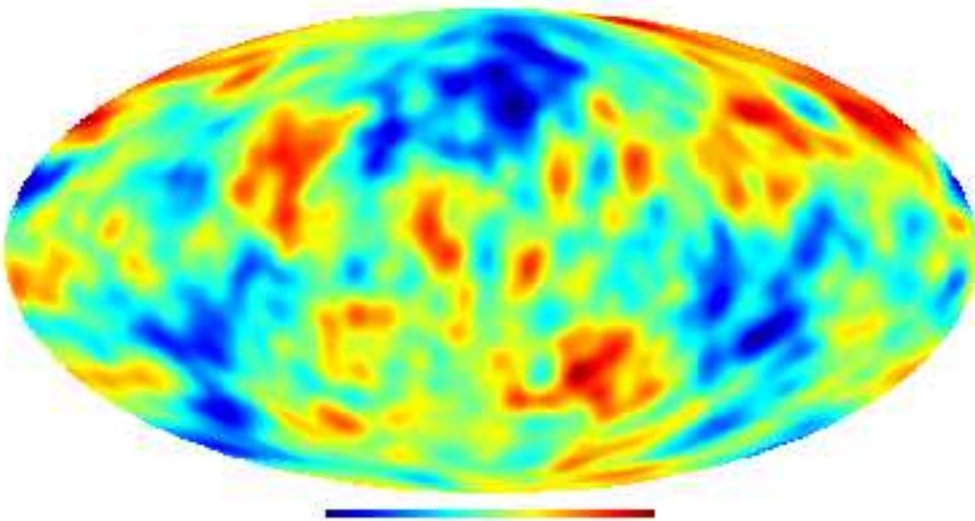,width=14 cm,angle=90}}
\caption{A non-Gaussian realisation at the same resolution generated
with the same $\Lambda$CDM power 
spectrum and a `white' bispectrum with $s=0.5$. The PDF $T(x)$ was
used as the non-Gaussian generator.}
\label{COBE1}
\end{figure}

\begin{figure}
\centerline{\psfig{file=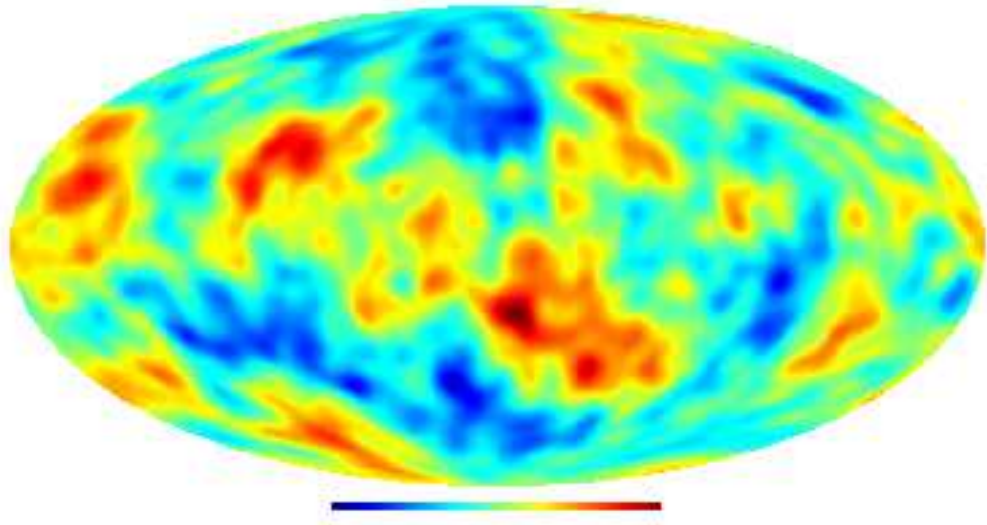,width=14 cm,angle=90}}
\caption{A similar realisation with $s=2.0$.}
\label{COBE2}
\end{figure}

\begin{figure}
\centerline{\psfig{file=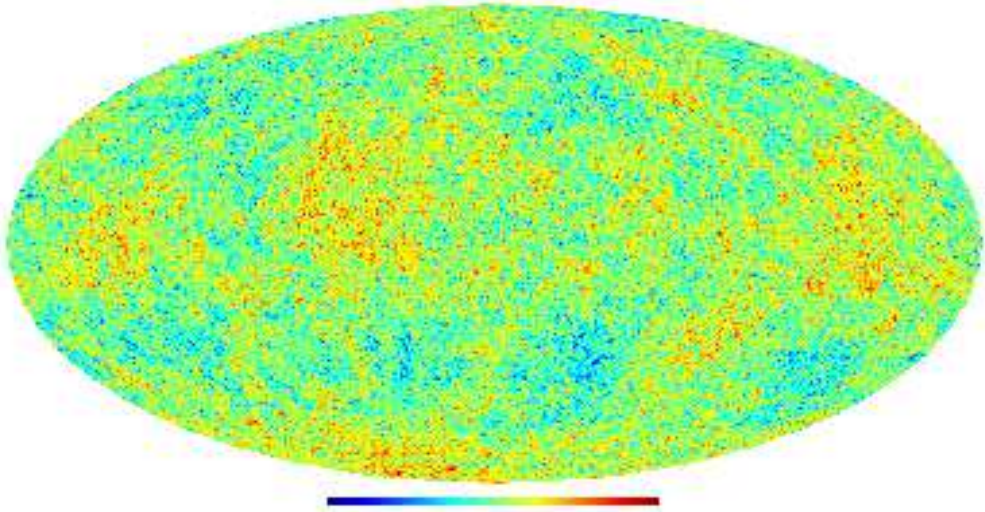,width=14 cm,angle=90}}
\caption{Same as above but at the resolution of the MAP experiment.}
\label{MAPng}
\end{figure}

In Fig.~\ref{COBE} we show a Gaussian realisation of a standard
$\Lambda$CDM  model at COBE-DMR \cite{COBE} resolution. The equivalent
non-Gaussian realisations with the same power spectrum are 
shown in Figs.~\ref{COBE1} and~\ref{COBE2} which are generated using
the PDF $T(x)$ and values for the parameter $s$ of $0.5$ and $2.0$
respectively.  Fig.~\ref{MAPng} shows a non-Gaussian realisation
($s=2.0$) with the same power spectrum and MAP \cite{MAP} resolution. 

We first make use of our maps to examine possible effects of non-Gaussianity
upon estimation of the power spectrum. For this purpose 
we compare the observed power
\be
C_{\ell}^{obs}=\frac{1}{2\ell+1}\sum_{m=-\ell}^{\ell}|a^{\ell 2}_m|, 
\ee
in the non-Gaussian ensembles with that 
of the usual Gaussian ensembles. Different PDFs will produce
ensembles with different cosmic variances (for the power) and in
particular the cosmic variance  will be different from that of
Gaussian ensembles. Assuming the $a_m^{\ell}$ modes to be Gaussian
distributed the power spectrum will have a $\chi_{2\ell +1}^2$
distribution which yields a particularly simple sample variance
\be
	\sigma^2(C^{obs}_{\ell}) = \frac{2C^2_{\ell}}{(2\ell+1)},
\ee
for zero noise and infinitely thin beam (see e.g. \cite{Knox95,Kendall}). 
In the non-Gaussian case the variance will assume a more complicated
form due to a non-zero contribution from the fourth order cumulant
(or connected moments in
analogy to the connected Green's functions of field theory):
\be
	\sigma^2(\mu_2) = \kappa_4+2\mu^2.
\ee 
In many motivated examples non-Gaussian processes display higher
cosmic variance in the power; an example is texture models as studied
in \cite{oldtext}.

\begin{figure}
\centerline{\hbox{\psfig{figure=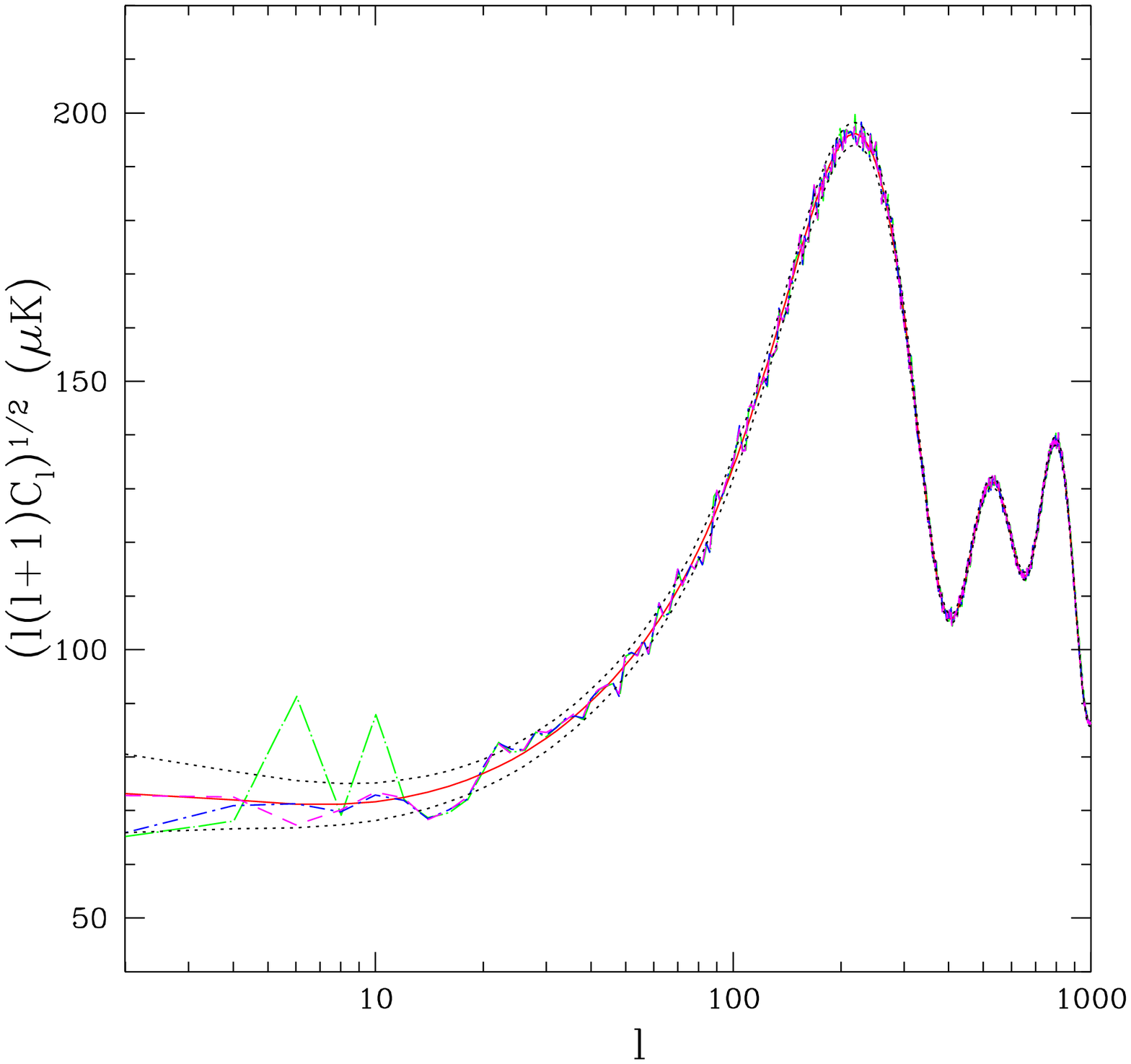,width=9cm}
\psfig{figure=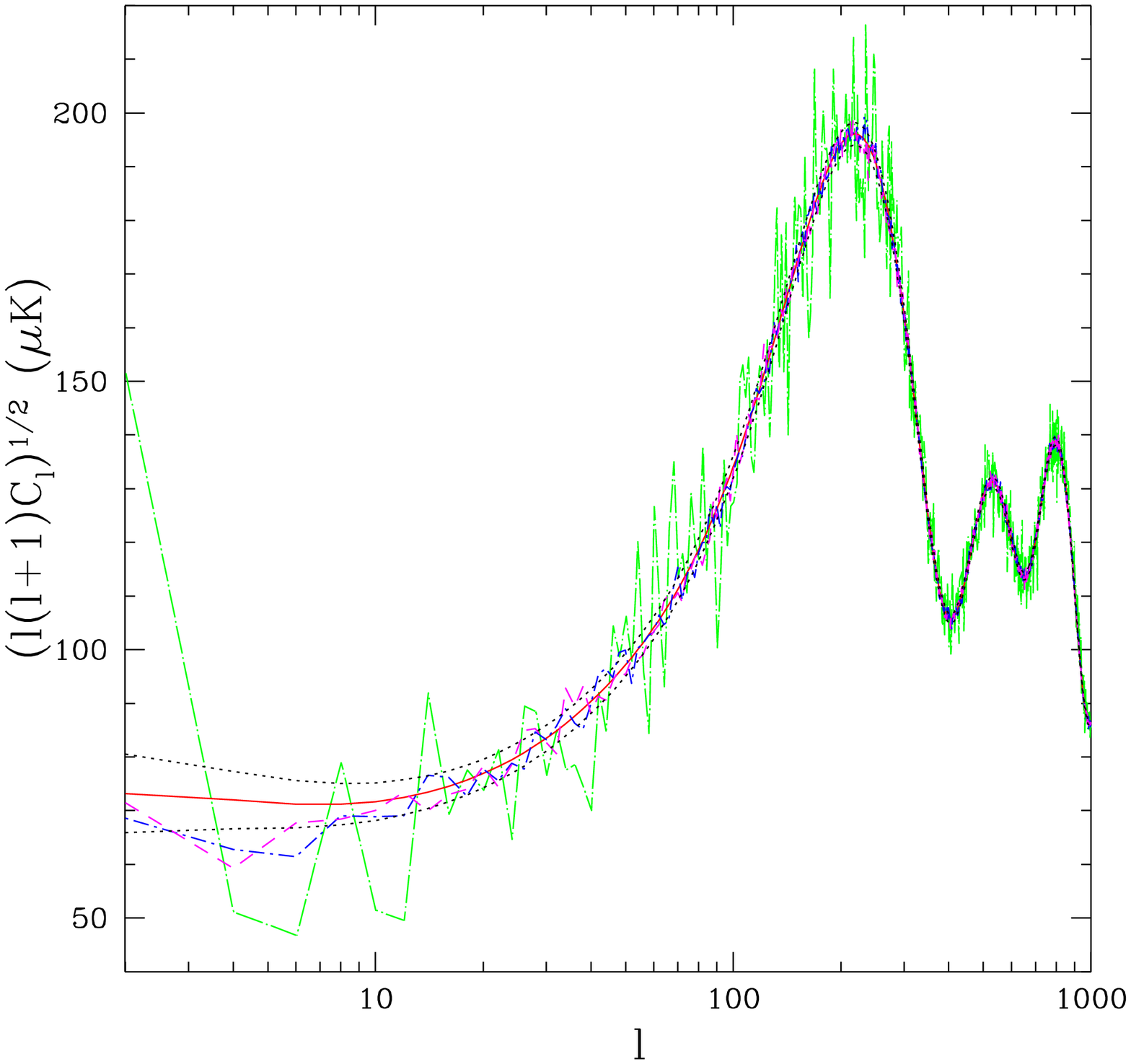,width=9cm} }}
\caption{The estimated power spectrum for ensembles of $10$ pure
signal maps. The solid line is the power spectrum of a standard
$\Lambda$CDM model ($\Omega_{\Lambda} = 0.7$, $\Omega_{CDM} = 0.25$,
$\Omega_{b} = 0.05$, $h=0.7$ and $n_s = 1.0$). The dotted lines show the extent
of cosmic variance of the averaged power for the respective  Gaussian
ensemble. The short dashed line is for an ensemble generated using the
distribution $P(x)$ with $s=0.5$. The short dashed dotted line is for
$s=0.5$ but generated using $T(x)$ and the long dashed dotted line is
for $T(x)$ with $s=5.0$. The ensembles in the left panel were
generated using only the $a_{\ell 0}$ modes as the non-Gaussian seeds in the
original anisotropic, uncorrelated ensembles whereas in the right
panel all the modes were originally non-Gaussian.}
\label{pow}
\end{figure}

In Fig.~\ref{pow} we show the average power spectrum for an ensemble
of $10$ maps for mild and extreme values for the relative skewness
parameter $s$. The maps do not include any noise contribution or beam
effects. The right panel shows the average power for maps where all
$a_{\ell m}$ modes are non-Gaussian in the original anisotropic,
uncorrelated ensemble. The left panel instead is for the case where
only the $a_{\ell 0}$ of the original modes are non-Gaussian. The
dotted line shows the extent of the $1$-sigma cosmic variance for an
equivalent Gaussian ensemble. 

We see that 
the sample variance of the power for the non-Gaussian variables
is comparable to that of its Gaussian counterpart for low $s$ and
grows for larger value of the relative skewness. This dependence is expected
since we know that $\kappa_4(\sigma,\al_0,\al_3)\equiv\kappa_4(s,C_{\ell})$.
Note that the isotropising rotation does not affect
ensemble averages, so that the sample variances of
estimators in the isotropic ensemble trace the
variances of the 1-D distributions generating the
non-Gaussianity. In general the contribution from the connected
moments can be derived analytically since they can be expanded in a
series of moments \cite{Kendall}. In this case though the solution for
$\sigma(\mu_2)$ in terms of $C_{\ell}$ is very contrived and has no
simple expression as the one above for the Gaussian case. Numerically
estimation on a case by case basis will give a more feasible
determination of the sample variance of the estimators being used.

\begin{figure}
\centerline{\hbox{\psfig{figure=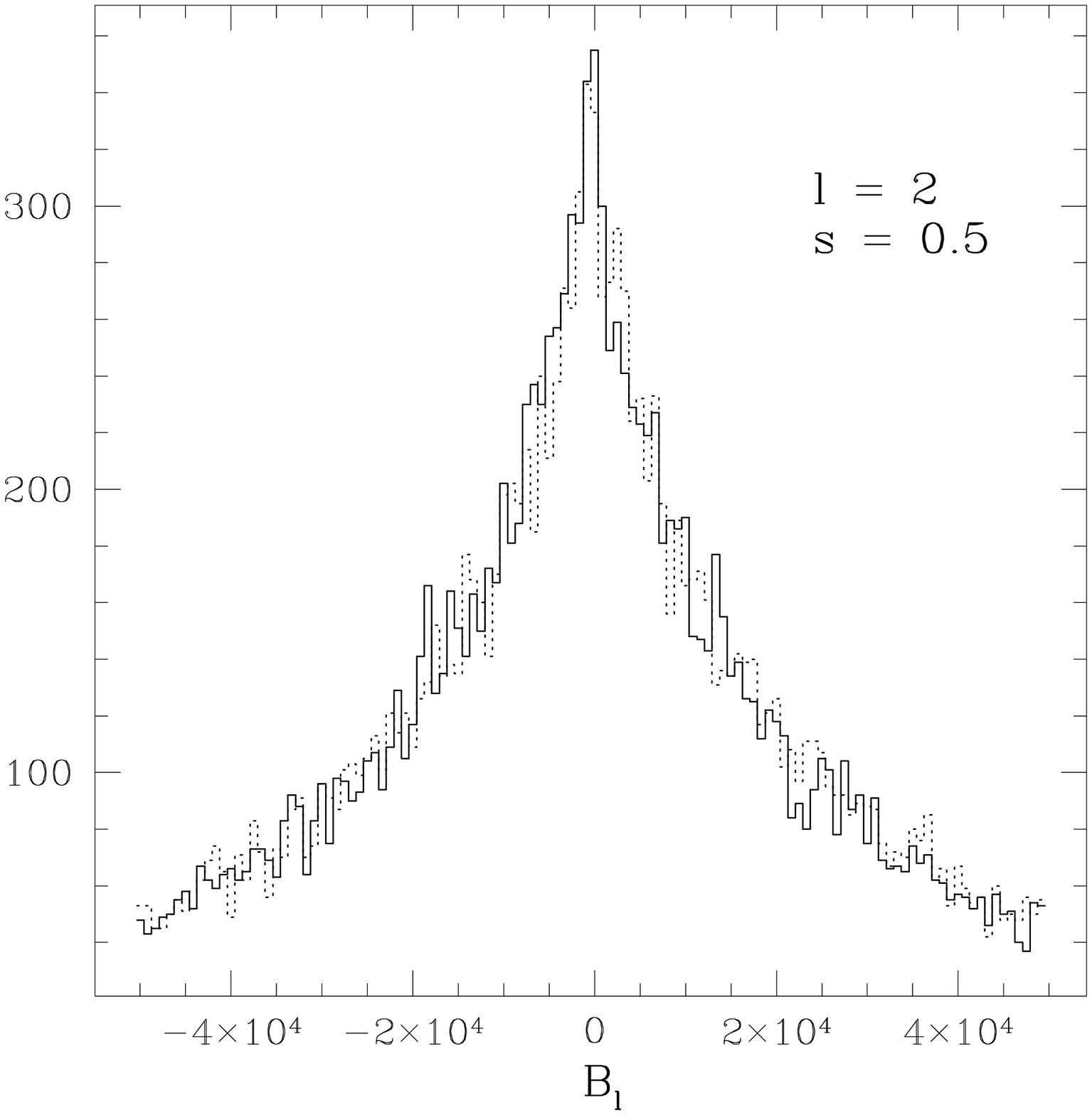,width=7cm}
\psfig{figure=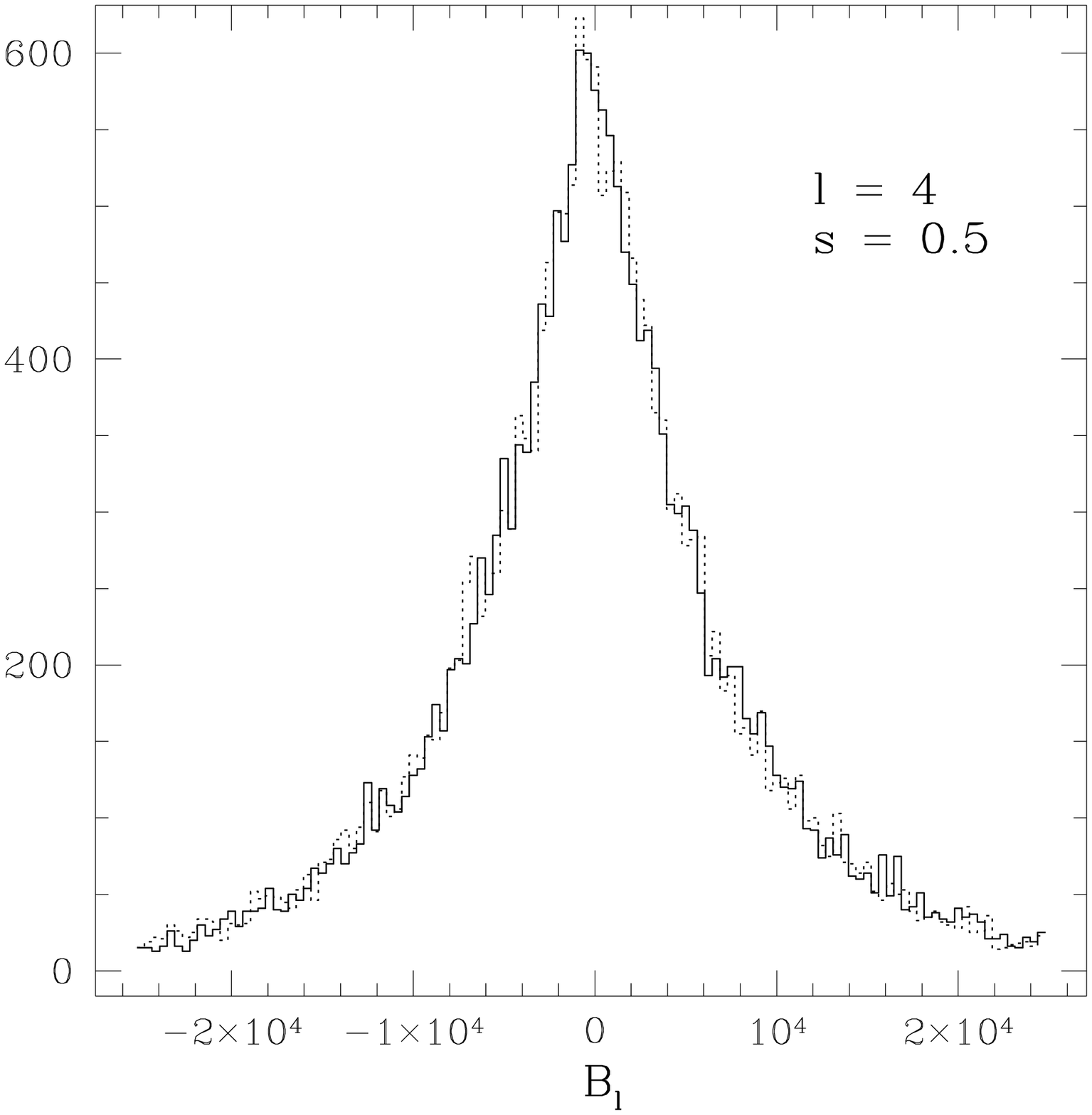,width=7cm}}}\hfill
\centerline{\hbox{ {\psfig{figure=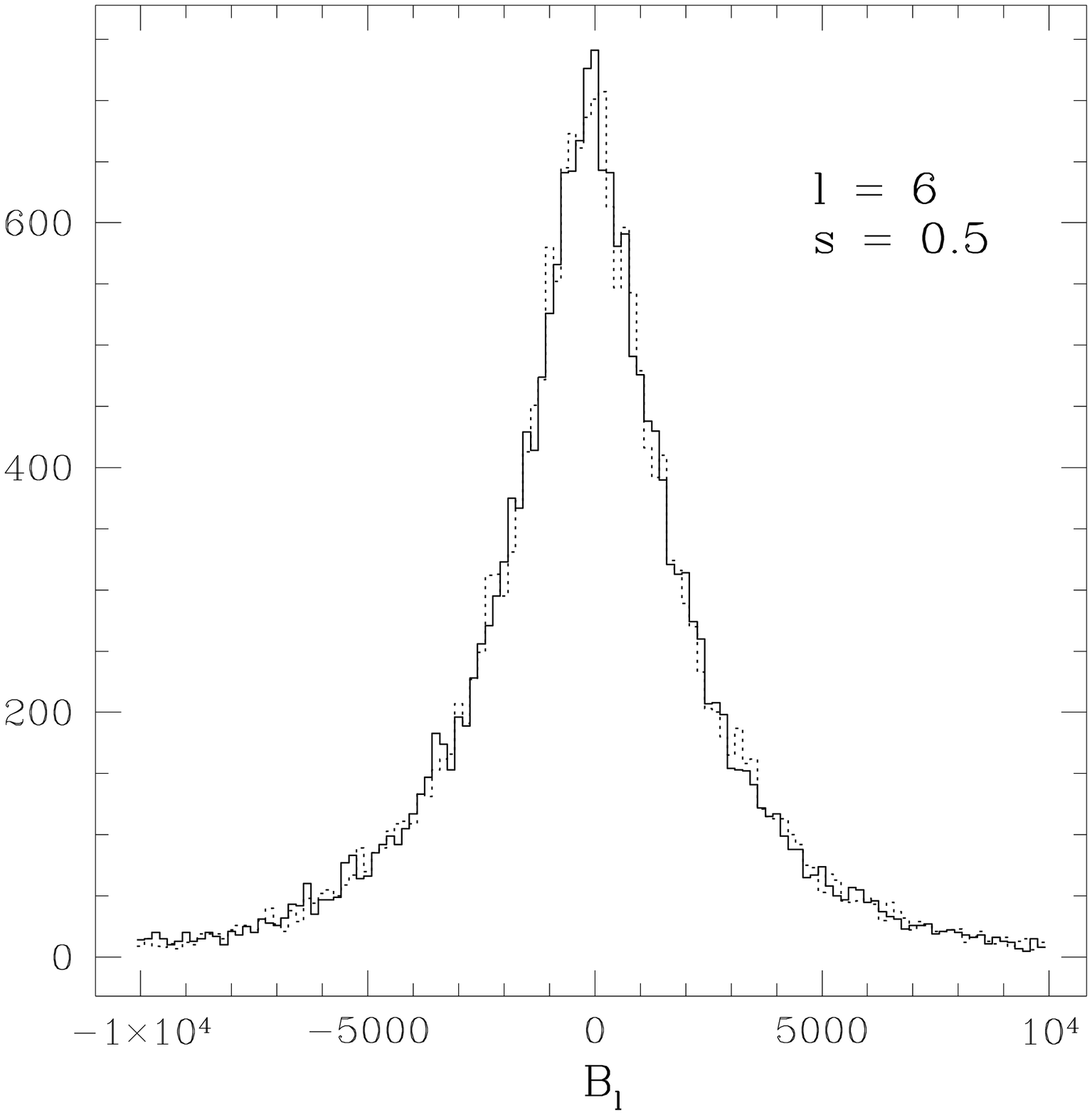,width=7cm}
\psfig{figure=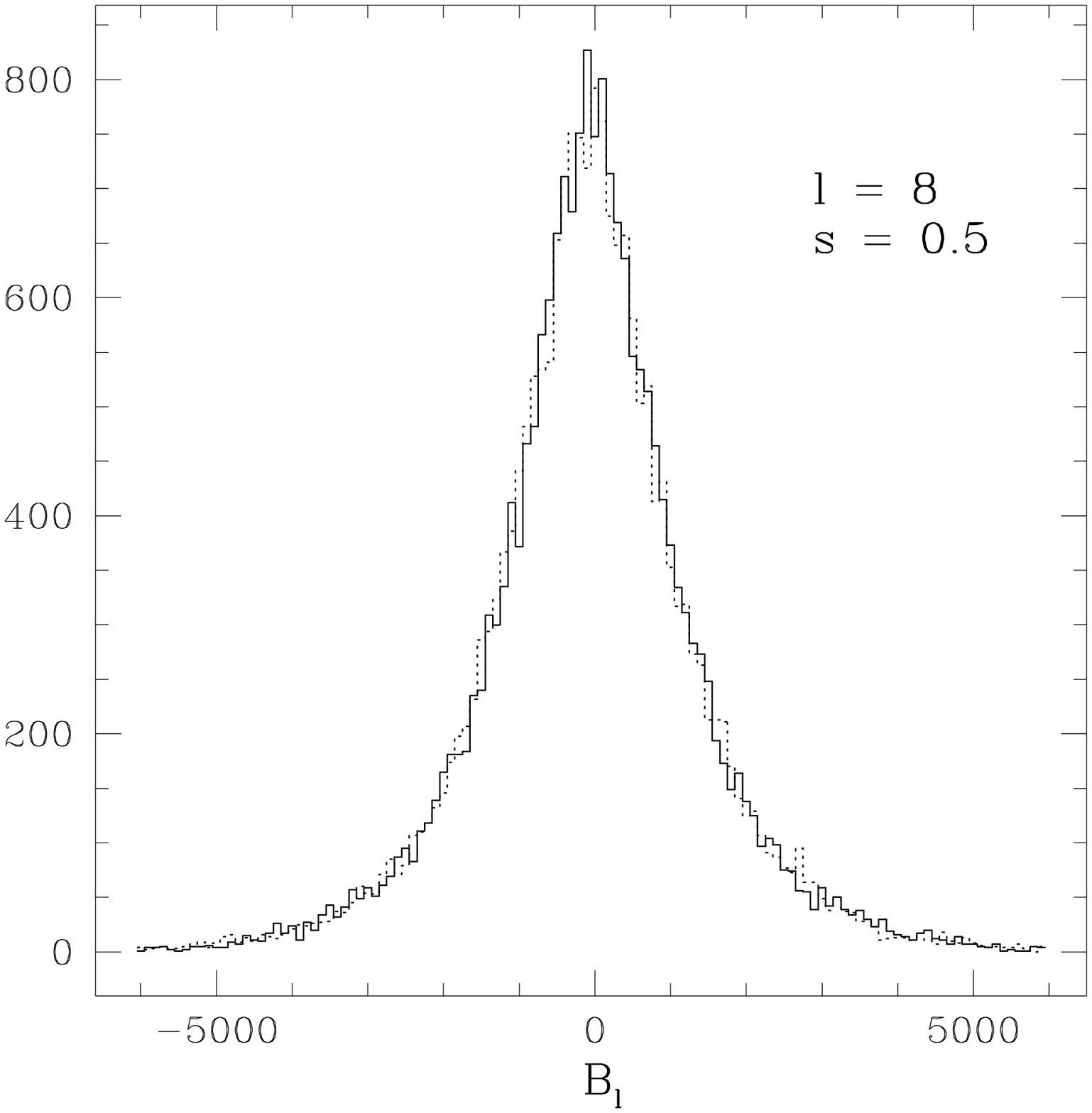,width=7cm} }}}
\caption{Histograms for the measured value of the estimator $B^{obs}_{\ell}$
for $20000$ 
realisations. The solid line corresponds to the non-Gaussian ensemble
generated using the distribution $T(x)$ with a value of $s=0.5$ and
the dashed line corresponds to the Gaussian ensemble.}
\label{tri}
\end{figure}

\begin{figure}
\centerline{\hbox{\psfig{figure=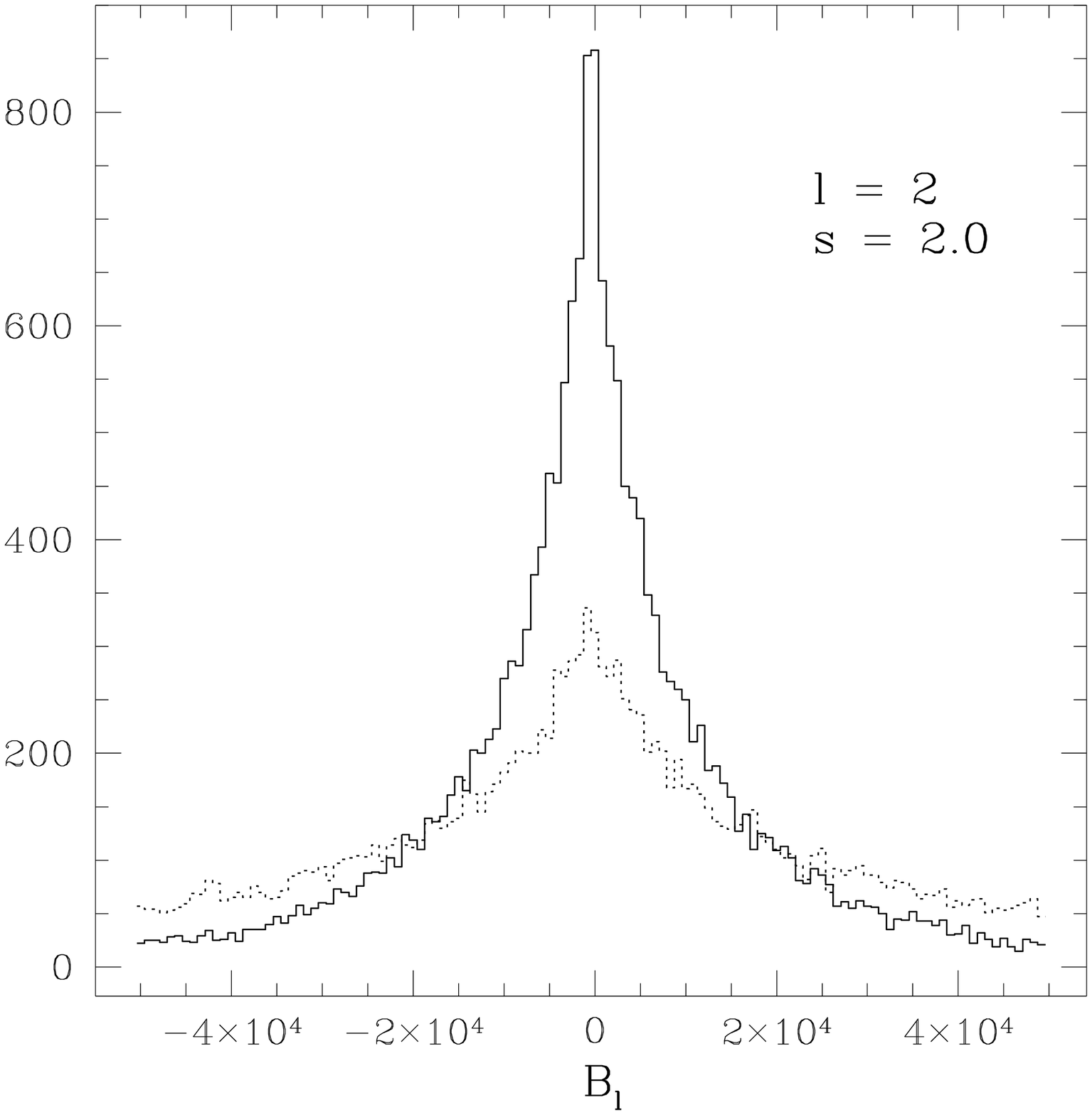,width=7cm}
\psfig{figure=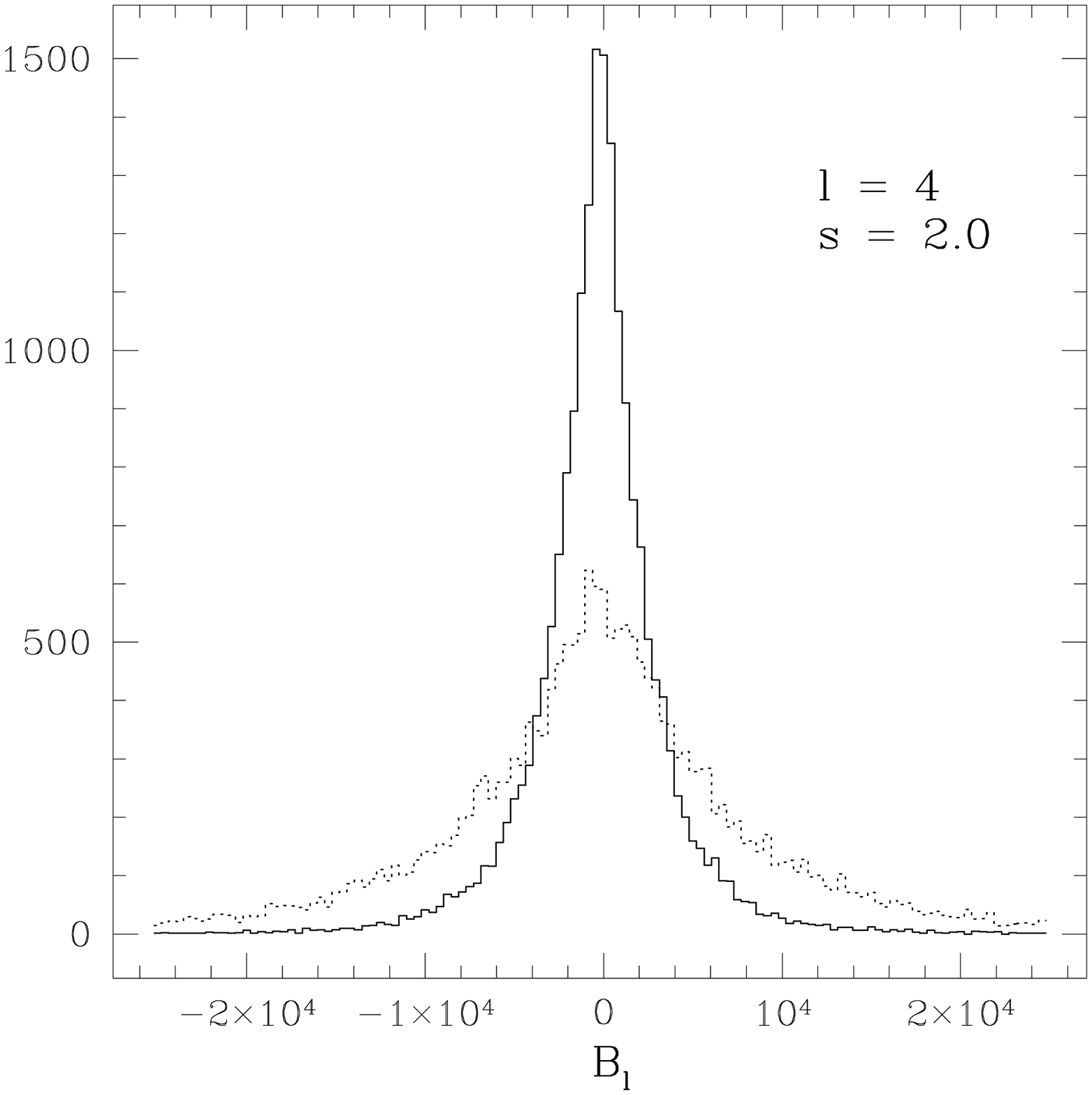,width=7cm}}}\hfill
\centerline{\hbox{ {\psfig{figure=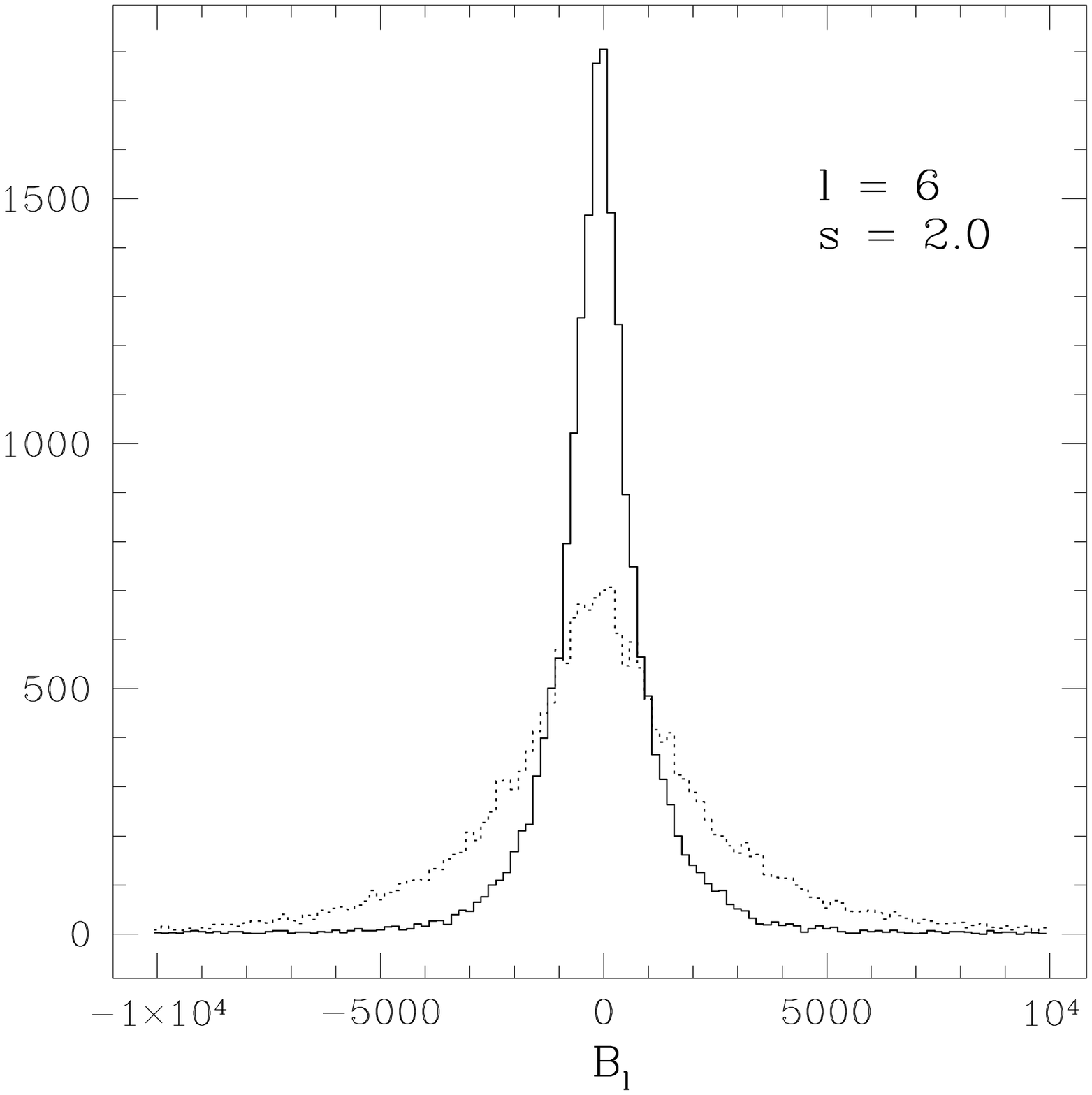,width=7cm}
\psfig{figure=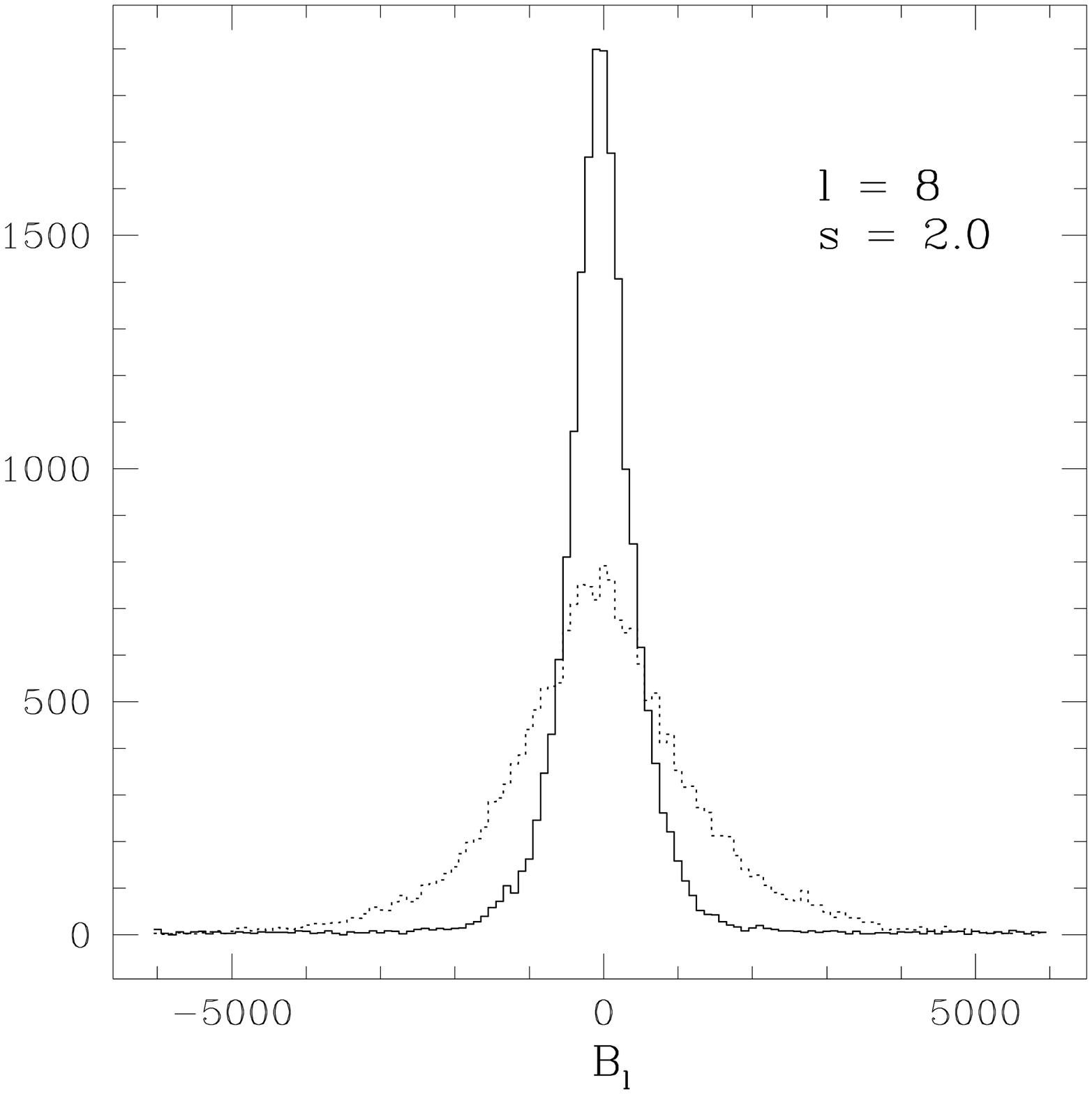,width=7cm} }}}
\caption{Similar to the previous plot but with $s=2.0$.}
\label{fig}
\end{figure}

We now turn to the issue of the detectability of non-Gaussian
bispectra making use of  our maps. 
In Figs.~\ref{tri} and~\ref{fig} we show the distribution of the
value for the observed bispectrum \cite{fmg}
\be
B^{obs}_{\ell}=\sum_{m_1m_2m_3}{\left( \begin{array}{ccc} \ell & \ell & \ell  \\ 
m_1 & m_2 & m_3
\end{array} \right )} a_{m_1}^{\ell}a_{m_3}^{\ell}a_{m_3}^{\ell},
\ee 
as measured from $20000$
realisation ensembles. 
Our Wigner coefficients are calculated using the recursive
relations of Schulten et al. \cite{schul} and are accurate to $\ell >
3000$.

We show the histograms for multipoles
$\ell=2,4,6$ and $8$ and for values of $s=2.0$ and $0.5$. We used the
distribution $T(x)$ to assign non-Gaussian values to all the $a_{\ell
m}$ modes. It is interesting to note that the sample variance of the
estimator $B_{\ell}^{obs}$ is reduced in the non-Gaussian case with
respect to that of the Gaussian ensemble. This shows that the
contribution from the connected moments to the non-Gaussian part of
$\mu_6$ is negative and in particular
\be
|\kappa_6| > 15\kappa_4\mu_2+9\kappa_3^2,
\ee 
confirming that the leakage into higher orders is not negligible for
both PDFs.

The implications of this result are interesting. It seems that,
for the type of non-Gaussianity which we have generated, 
when searching for non-Gaussianity and armed with a single 
realisation, we could only rely on the {\it rejection} of the non-Gaussian
hypothesis by observing modes whose bispectra are
(counter-intuitively) too large for them to be non-Gaussian. 
Conversely if we were to measure bispectra which are consistently 
too close to zero for enough $\ell$ we could reject Gaussianity
in favour of non-Gaussianity. Even though this remark may
sound counter-intuitive it is the basis of the result in 
\cite{mag1} (in which Gaussianity was rejected on the basis of
a low $\chi^2$).

\begin{table}[ht]
\begin{center}
\begin{tabular}{c|ccc|ccc}
$\ell$ & $ C_{\ell}$ & $(C_{\ell}^{obs})_{G}$ &
$(C_{\ell}^{obs})_{NG}$ & $B_{\ell}$ & $(B_{\ell}^{obs})_{G}$ & 
$(B_{\ell}^{obs})_{NG}$ \\
\hline
  2 & 892.729 & 890.716 & 892.590 & -4938.244 & -325.169  & -6376.174\\
  4 & 259.111 & 258.795 & 257.526 & 426.451 &    -97.583  & 559.303\\
  6 & 120.751 & 120.684 & 121.219 & -160.554 &    -4.521  & -123.481\\
  8 &  70.419 & 70.401 & 70.388  &  13.399 &     -11.157  & 42.092\\
 10 &  46.657 & 46.656 & 46.687  & -27.442 &       3.777  & -18.385\\
 12 &  33.704 & 33.793 & 33.783  &   5.682 &      -8.440  & 9.484\\
 14 &  25.713 & 25.674 & 25.763  &  -6.365 &      -5.433  & -5.449\\
 16 &  20.419 & 20.480 & 20.383  &   1.775 &       0.546  & 3.389\\
 18 &  16.747 & 16.737 & 16.730  &  -4.908 &      -0.757  & -2.245\\
 20 &  14.084 & 14.087 & 14.077  &   2.081 &       0.431  & 1.563\\
\end{tabular}
\end{center}
\caption{The observed statistics for the first 10 even multipoles for
both a Gaussian and non-Gaussian noisless ensemble.}
\label{tab1}
\end{table}

For completeness we have displayed in Table~\ref{tab1} all the results
concerning $C_{\ell}^{obs}$ and $B_{\ell}^{obs}$ in
both the Gaussian and non-Gaussian ensembles as discussed in the previous
paragraphs.

\section{Discussion}\label{disc}
In this paper we proposed a method with which to generate
non-Gaussian maps with fixed power spectrum and bispectrum. 
Our strategy is as follows.
We generate these maps in harmonic space, and give each of the
modes, independently,  
a 1-D non-Gaussian PDF. The resulting maps are anisotropic,
but a random rotation then restores the ensemble isotropy. 
We proved that the power spectrum and bispectrum of the isotropic
ensemble is simply related to the variance and skewness of the 
1-D PDFs employed. We then
proposed two different PDFs for which the variance and skewness
may be fixed independently. One is based upon the Hilbert space
of an harmonic oscillator; the other upon a superposition of Gaussian
functions. In both cases we worked out the algebra relating the
parameters controlling the distributions and the variance and
skewness. We then drew our random numbers numerically using a
rejection method. 

Even though we only considered the generation of maps with
a fixed power spectrum and bispectrum
we stress that the  extension of our method for higher order
moments is straightforward, even though it has its computational
costs. The various parameters of the PDFs considered may be used
to fix the higher order cumulants of the mother 1-D PDF applied
to single modes. Once again a random rotation restores isotropy,
and the isotropic higher-order correlators may be simply related
to the cumulants of the original PDFs. 

We close with an evaluation of the efficiency of our method when
meeting the ever improving resolutions of upcoming experiments.
As we have shown it is possible to generate high resolution
non-Gaussian maps with this method. However they are computationally
more expensive to produce than Gaussian maps, specifically due
to the random rotation to which they must be subject. Currently the
rotation is carried out in $\ell$-space which requires the computation of
the rotation matrix elements \cite{Var}. This avoids sampling problems
which would arise if it were carried out in pixel but requires
increasingly long computation times for increasing $\ell$.  
On the other hand the real bottleneck is likely to be at the analysis
stage, rather than in the simulation of maps. Sums involving Wigner
coefficients become computationally intensive as $\ell$ is
increased. 

\section*{Acknowledgments}
We would like to thank P. Ferreira and G. Rocha for useful
discussions.

\end{document}